\documentclass[11pt,a4paper,nofootinbib,superscriptaddress]{revtex4-1}

\pdfoutput=1
\usepackage[utf8]{inputenc}
\usepackage{graphicx,bm,bbm}
\usepackage{amssymb,graphicx,epstopdf}
\usepackage{slashed,subfigure}
\usepackage{caption}
\usepackage{xcolor}
\usepackage{amsmath,mathtools}
\usepackage{epstopdf,dcolumn}
\usepackage{soul}
\usepackage{mathtools}
\allowdisplaybreaks
\usepackage[normalem]{ulem}
\usepackage{cancel}
\usepackage{xcolor}
\usepackage[colorlinks=true,linktocpage=true,linkcolor=blue,citecolor=blue]{hyperref}
\usepackage{xcolor}
\usepackage{braket}
\allowdisplaybreaks
\usepackage{enumitem}
\usepackage{mathrsfs}

\def\be {\begin{equation}}
\def\ee {\end{equation}}
\def\bea {\begin{eqnarray}}
\def\eea {\end{eqnarray}}
\def\bc {\begin{center}}
	\def\ec {\end{center}}
\def\nn {\nonumber}
\def\eps {\epsilon}
\def\gm {\gamma}

\def\al {\alpha}

\def\({\left(}
\def\){\right)}
\def\[{\left[}
\def\]{\right]}

\newcommand \Tr{\operatorname{\text{Tr}}}

\def\slashed{\slash\!\!\!\!}

\DeclareGraphicsExtensions{.jpg,.pdf,.eps}

\begin{document}

	\title{A study of shear viscosity coefficient of quark matter in Polyakov quark meson model for three flavors}
	
	\author{Ritesh Ghosh}
	\email{ritesh.ghosh@saha.ac.in}
	\affiliation{
		Theory Division, Saha Institute of Nuclear Physics, HBNI, \\
		1/AF, Bidhannagar, Kolkata 700064, India}
	\affiliation{
		Homi Bhabha National Institute, Anushaktinagar, \\
		Mumbai, Maharashtra 400094, India}

	\begin{abstract}
	We have computed shear viscosity coefficient for hot and dense QCD matter within Polyakov loop extended quark meson(PQM) model for three flavors. Viscosity coefficient is estimated using Boltzmann equation within the relaxation time approximation. Relaxation times for quarks are calculated from elastic scattering of quarks and antiquarks via gluon exchange. Also the mean free paths of strange and light quarks are estimated for this cases. The calculations are performed at finite chemical potential. Thermodynamic properties like energy, pressure, entropy densities are also computed and compared with lattice QCD data.
	\end{abstract}
	
	\maketitle 
	\newpage
	
\section{Introduction}
A state of matter of quarks and gluons i.e. quark-gluon plasma (QGP) is formed in relativistic heavy ion collisions. Different signatures like jet quenching~\cite{Cao:2020wlm}, elliptic flow~\cite{Csernai:1999nf}, quarkonia suppression~\cite{Brambilla:2010cs}, photon/dilepton production~\cite{Rapp:2002tw, Shuryak:1992bt} etc from high energy heavy-ion collision experiments at Relativistic Heavy-Ion Collider (RHIC) and at Large Hadron Collider (LHC) suggest the creation of such deconfined QGP state. The produced hot and dense deconfined matter expands and cools down and then undergoes through a phase transition to the hadronic final state. Transport coefficients are interesting quantities in this context. The temperature and chemical potential dependency of transport coefficients draws relevant impact on the location of phase transition.  Elliptic flow~\cite{Romatschke:2007mq, Gale:2013da} study describes the collectivity of  the hot and dense quark-gluon matter at high temperature and indicates the smallest viscosity to entropy density ratio $(\eta/s)$. These studies suggest that the QGP produced in heavy-ion collisions interacts strongly and behaves as a nearly perfect relativistic fluid~\cite{Schafer:2009dj}. For hydrodynamic simulations, the transport coefficients are also input parameters to explain the evolution of the hot and dense matter~\cite{Csernai:2006zz, Heinz:2013th}.

Transport coefficients can be computed from the QCD using Kubo formulas~\cite{Kubo:1957mj, Akemann:2002js, Lang:2012tt}. But it is complicated to calculate from first principles. Lattice QCD~\cite{Borsanyi:2013bia, Nakamura:2004sy} may be a good approach to calculate transport coefficient at zero chemical potential. But it is very challenging to deal with finite chemical potential due to the well known fermion sign problem for QCD. Hence, at higher chemical potential  several investigations have been done to calculate shear viscosity of strongly interacting matter using effective model calculations like Nambu-Jona- Lasinio (NJL)~\cite{Marty:2013ita, Deb:2016myz}  , quasiparticle model (QPM )~\cite{Sasaki:2008fg, Bluhm:2010qf, Soloveva:2019xph}, chiral perturbation theory~\cite{Dobado:2011qu, Itakura:2007mx}, the linear sigma model~\cite{Chakraborty:2010fr, Heffernan:2020zcf}, the functional
renormalization group approach~\cite{Pawlowski:2005xe, Bagnuls:2000ae, Gies:2006wv}. To explore the QGP phase transition experimentally at medium and higher chemical potential, the Beam Energy Scan (BES) program at RHIC~\cite{Odyniec:2019kfh} has carried out several experiments. The near future experimental program of FAIR (Facility for Antiproton and Ion Research)~\cite{Senger:2020fvj} at GSI and the NICA(Nuclotron-based Ion Collider Facility) facility at Joint Institute
for Nuclear Research (JINR)~\cite{Sissakian:2009zza} are also likely to explore more physics at moderate and relatively high chemical potential.
 
Effective models like NJL, quark meson (QM) model can explain the QCD chiral phase diagram. But they can not address the confinement-deconfinement transition scenario. Both the picture can be achieved by incorporating Polyakov loop in NJL or QM type models. Polyakov NJL (PNJL)~\cite{Ciminale:2007sr, Fukushima:2008wg, Fu:2007xc} or Polyakov quark meson (PQM)~\cite{Mao:2009aq, Stiele:2016cfs} models for three flavors can agree better with lattice data.
In the present article we have investigated the shear viscosity coefficient within the Polyakov loop extended quark meson model for $(2+1)$ flavors at finite temperature and chemical potential. Shear viscosity coefficient is calculated within the relaxation time approximation of the Boltzmann equation. Medium dependent particle masses are used from the PQM model. The relaxation time is estimated from the scattering rate of quark quark and quark antiquark via gluon exchange. In Ref.~\cite{Abhishek:2017pkp, Singha:2017jmq}, the authors calculated transport coefficients for quarks and hadronic matters within two flavor PQM model.

The article is organized as follows. In Sec.~\ref{pqm_model}, the formalism of three flavor Polyakov loop quark meson model is addressed. We get the medium dependent masses of particles which are involved in calculation of transport coefficients. Thermodynamic properties like energy, pressure, entropy densities are also discussed in Sec.~\ref{thermodynamic}. Relaxation time and mean free path is obtained by studying the quark-quark and quark-antiquark scattering for three quark flavors. We have derived shear viscosity in Sec.~\ref{transport}.
Numerical results are described in Sec.~\ref{result} and finally the work is  summarized in Sec.~\ref{summary}.
\section{Polyakov Quark-meson model}
\label{pqm_model}
\subsection{Polyakov loop}
Three flavor PQM model~\cite{Schaefer:2009ui} is a generalization of two flavor PQM model~\cite{Schaefer:2007pw} combining the chiral symmetry restoration and the feature of confinement-deconfinement transition.
Polyakov loop operator is defined as Wilson loop in the temporal direction~\cite{POLYAKOV1978477}
\bea
\mathcal{P}(\vec{x})&=& P \exp\Big(i \int_0^\beta d\tau A_0(\vec{x},\tau)\Big),
\eea
where $P$ is the path ordering and $A_0$ is the temporal component of Euclidean gauge field $A_\mu$.
Polyakov loop variables are 
\bea
\Phi(\vec{x}) &=& \frac{1}{N_c}\langle\Tr \mathcal{P}(\vec{x})\rangle,\\
\bar\Phi(\vec{x}) &=& \frac{1}{N_c}\langle\Tr \mathcal{P}(\vec{x})^\dagger\rangle,
\eea
where trace is taken over color space. $N_c$ is color degrees of freedom. Mean value of $\Phi$ and $\bar \Phi$ are related to the free energy of infinitely heavy, static quarks and antiquarks.  $\Phi$ and $\bar \Phi$ vanish in the confined phase as infinite energy is required to put a static quark in that phase.  $\Phi$ and $\bar \Phi$ are finite in deconfined phase. So $\Phi$ works as the order parameter of confinement-deconfinement phase transition.

The PQM Lagrangian contains quark-meson contribution and Polyakov loop potential.
The form of PQM Lagrangian reads as~\cite{Schaefer:2009ui}
\bea
L_{PQM}&=& \overline\psi (i \,\slashed D-g \phi_5)\psi +L_m -U(\phi, \overline \phi),
\eea
where $\psi\equiv (u,d,s)$ indicates quark filed for $N_f=3$ flavors. Covariant derivative is defined as $D_\mu=\partial_\mu-i A_\mu$ with $A_\mu=\delta_{\mu 0}A^0$.  Here $A_\mu=g_s A_\mu^a \lambda^a/2$, where $\lambda^a$ with $a=1,..., N_c^2-1$ are usual Gell-Mann matrices and $g_s$ is the $SU(N)$ gauge coupling.  Quark meson interaction is incorporated by the flavor blind Yukawa coupling $g$ and the meson matrix
\bea
\phi_5=T_a (\sigma_a+i \gm_5 \pi_a).
\eea
$\sigma_a$ and $\pi_a$ are nine scalar and pseudoscalar mesons respectively. 
$T_a=\lambda_a/2$ represent the nine generators of the $U(3)$ symmetry. The generators obey the usual $U(3)$ algebra. 

Pure mesonic contribution can be written as
\bea
L_m&=& \Tr(\partial_\mu \phi^\dagger \partial ^\mu \phi)-m^2 \Tr(\phi^\dagger \phi)-\lambda_1[ \Tr(\phi^\dagger \phi)]^2-\lambda_2  \Tr(\phi^\dagger \phi)^2\nn\\
&+& c[\det(\phi)+\det(\phi^\dagger)]+\Tr[H (\phi+\phi^\dagger)],
\eea
where $\phi$ is a $3\times 3$ matrix defined as
\bea
\phi= T_a \phi_a = T_a (\sigma_a +i \pi_a).
\eea
Chiral symmetry is explicitly broken by the term $H=T_a h_a$, where $h_a$ are nine external parameters. $m$ is the tree level mass of meson fields. $\lambda_1$ and $\lambda_2$ are two quartic coupling constant and $c$ is cubic coupling constant which models the $U_A(1)$ axial anomaly of the QCD vacuum. 

\subsection{Potential for the Polyakov loop}
Polyakov loop potential for $SU(3)$ pure gauge theory has been proposed to reproduce the lattice thermodynamic results. There are several forms of the potential in literature. Here we adopt the polynomial parametrization form of the potential as~\cite{Ratti:2005jh,Schaefer:2009ui}
\bea
U_p(\Phi,\bar \Phi, T)=T^4\bigg[-\frac{b_2}{4}\big(|\Phi|^2+|\bar \Phi|^2\big)-\frac{b_3}{6}\big(\Phi^3+\bar \Phi^3\big)+\frac{b_4}{16}\big(|\Phi|^2+|\bar \Phi|^2\big)^2\bigg],
\eea
with the temperature dependent coefficient
\bea
b_2(T)=a_0+a_1\bigg(\frac{T_0}{T}\bigg)+a_2\bigg(\frac{T_0}{T}\bigg)^2+a_3\bigg(\frac{T_0}{T}\bigg)^3,
\eea
governing the confinement-deconfinement phase transition. At low temperature, the potential $U_p(\Phi,\bar \Phi, T)$ has one minimum at $\Phi=0$ in confined phase. Above the critical temperature, due to the spontaneous breaking of $Z(3)$ central symmetry, the potential has three degenerate global minima at non zero $\Phi$. Here we mention the value of the temperature independent parameters:
\bea
a_0=6.75, \,\, a_1=-1.95,\,\, a_2=2.625,\,\, a_3=-7.44,\,\, b_3=0.75,\,\, b_4=7.5 .\nn
\eea

\subsection{Thermodynamic Potential}
To investigate the thermodynamic properties of the PQM model the thermodynamic potential is used in light of mean field approximation. Consisting the mesonic $ \mathcal{U}(\sigma_x,\sigma_y)$, quark-antiquark $\Omega_{q\overline q}(\Phi,\bar \Phi )$ and Polyakov loop contribution $U_p(\Phi,\bar \Phi )$, the thermodynamic potential can be given by,
\bea
\Omega(T,\mu) &=& \mathcal{U}(\sigma_x,\sigma_y)+U_p(\Phi,\bar \Phi )+\Omega_{\overline q q}(\Phi,\bar \Phi ,\sigma_x,\sigma_y).
\eea
The mesonic contribution reads as~\cite{Tawfik:2014gga, Schaefer:2008hk, Gupta:2009fg},
\bea
\mathcal{U}(\sigma_x,\sigma_y)&=&\frac{m^2}{2}(\sigma_x^2+\sigma_y^2)-h_x\sigma_x-h_y\sigma_y-\frac{c}{2\sqrt{2}\sigma_x^2\sigma_y}+\frac{\lambda_1}{2}\sigma_x^2\sigma_y^2\nn\\
&+&\frac{1}{8}(2\lambda_1+\lambda_2)\sigma_x^4+\frac{1}{8}(\lambda_1+2\lambda_2)\sigma_y^4.
\eea

Minimizing the grand potential with respect to $\sigma_x$,$\sigma_y$,$\Phi$ and $\bar \Phi$, we obtain four coupled equations. There are six input parameters $ m^2, c,\lambda_1, \lambda_2, h_x, h_y$ and they are fitted with known quantities: pseudoscalar masses i.e. kaon mass $m_k$, pion mass $m_\pi$, $m_\eta^2+m_{\eta'}^2$ , pion and kaon decay constant $f_\pi$, $f_K$ and mass of scalar $\sigma$ meson $m_\sigma$. These parameters are fitted to produce the observed pion mass in vacuum.

The fermionic part of the thermodynamic potential is written as
\bea
\Omega_{\overline qq}(\Phi,\bar \Phi ,\sigma_x,\sigma_y)&=&-2T \sum_{f=u,d,s}\int \frac{d^3p}{(2\pi)^3}\bigg\{\ln g_f^+ + \ln g_f^-\bigg\},
\eea
where  $g_f^+$ and $g_f^-$ are defined as
\bea
g_f^+&=&\bigg[1+3(\Phi+\bar \Phi e^{-(E_f-\mu_f)/T})e^{-(E_f-\mu_f)/T}+e^{-3(E_f-\mu_f)/T}\bigg]\\
g_f^-&=&\bigg[1+3(\bar \Phi+ \Phi e^{-(E_f+\mu_f)/T})e^{-(E_f+\mu_f)/T}+e^{-3(E_f+\mu_f)/T}\bigg].
\eea
where $E_f=\sqrt{p^2+m_f^2}$ is the flavor dependent single particle energy with the quark masses 
\bea
m_{u,d}=g\sigma_x/2 ,\,\,\, m_s=g \sigma_y/\sqrt{2}\,.
\label{quark_mass}
\eea
$\mu_f$ is the chemical potential for a quark flavor $f$. In the present work, uniform quark chemical potential $\mu=\mu_u=\mu_d=\mu_s$ is considered. 
Finally, the behavior of quark condensate $\sigma_x, \sigma_y$ and the Polyakov loop expectation values $\Phi, \bar \Phi$ can be obtained by minimizing the total effective potential $\Omega$ for a given value of temperature and chemical potential i.e.
\bea
\frac{\partial \Omega}{\partial\sigma_x}=\frac{\partial \Omega}{\partial\sigma_y}=\frac{\partial \Omega}{\partial\Phi}=\frac{\partial \Omega}{\partial\bar\Phi}\Bigg|_{\sigma_x=\langle\sigma_x\rangle,\sigma_y=\langle\sigma_y\rangle,\Phi=\langle \Phi\rangle, \bar\Phi =\langle \bar\Phi\rangle}=0.
\label{gap_eq}
\eea

 \section{QCD Thermodynamics}
 \label{thermodynamic}
 
  \begin{center}
 	\begin{figure}[tbh]
 		\begin{center}
 			\includegraphics[scale=0.5]{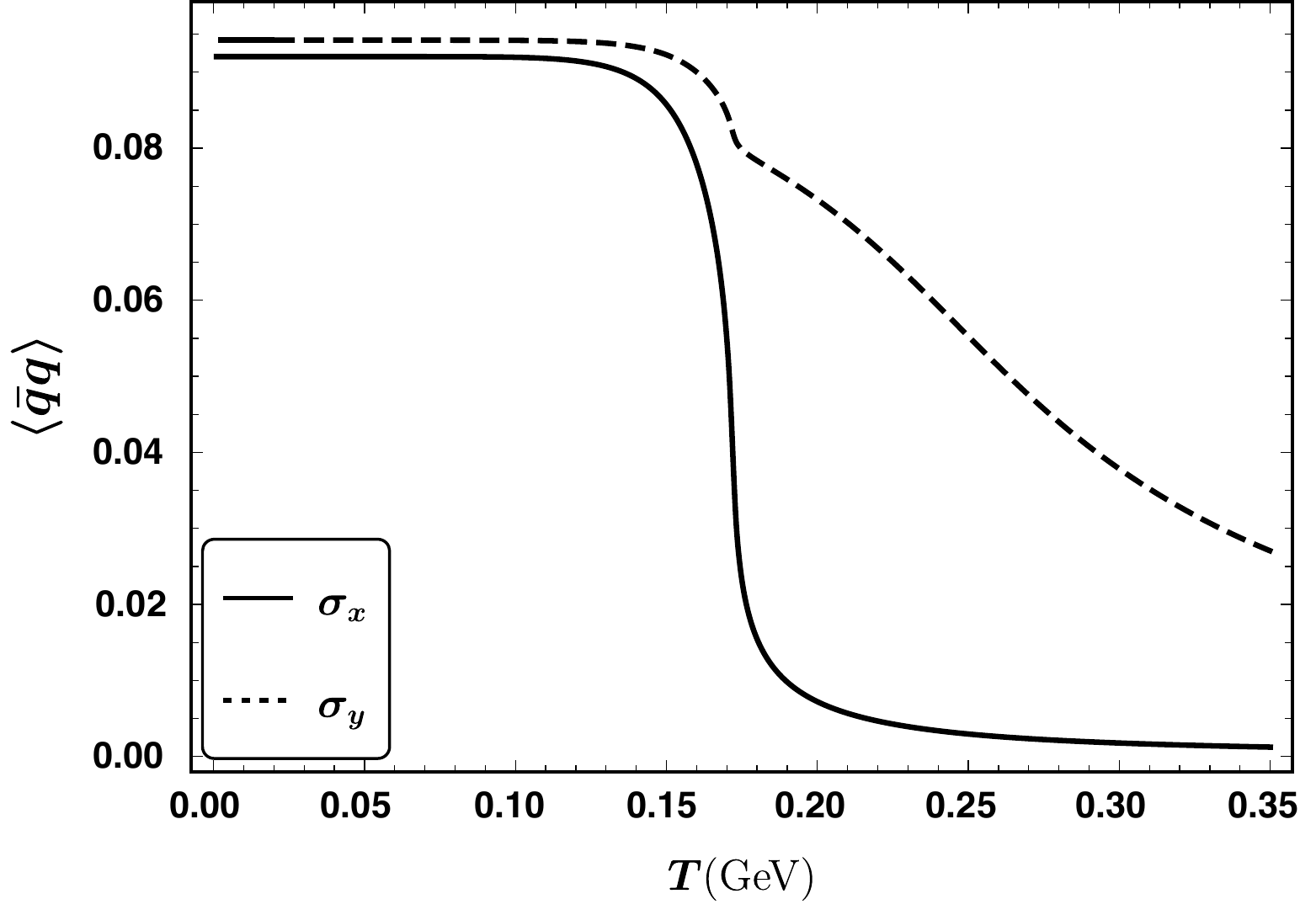}
 			\includegraphics[scale=0.5]{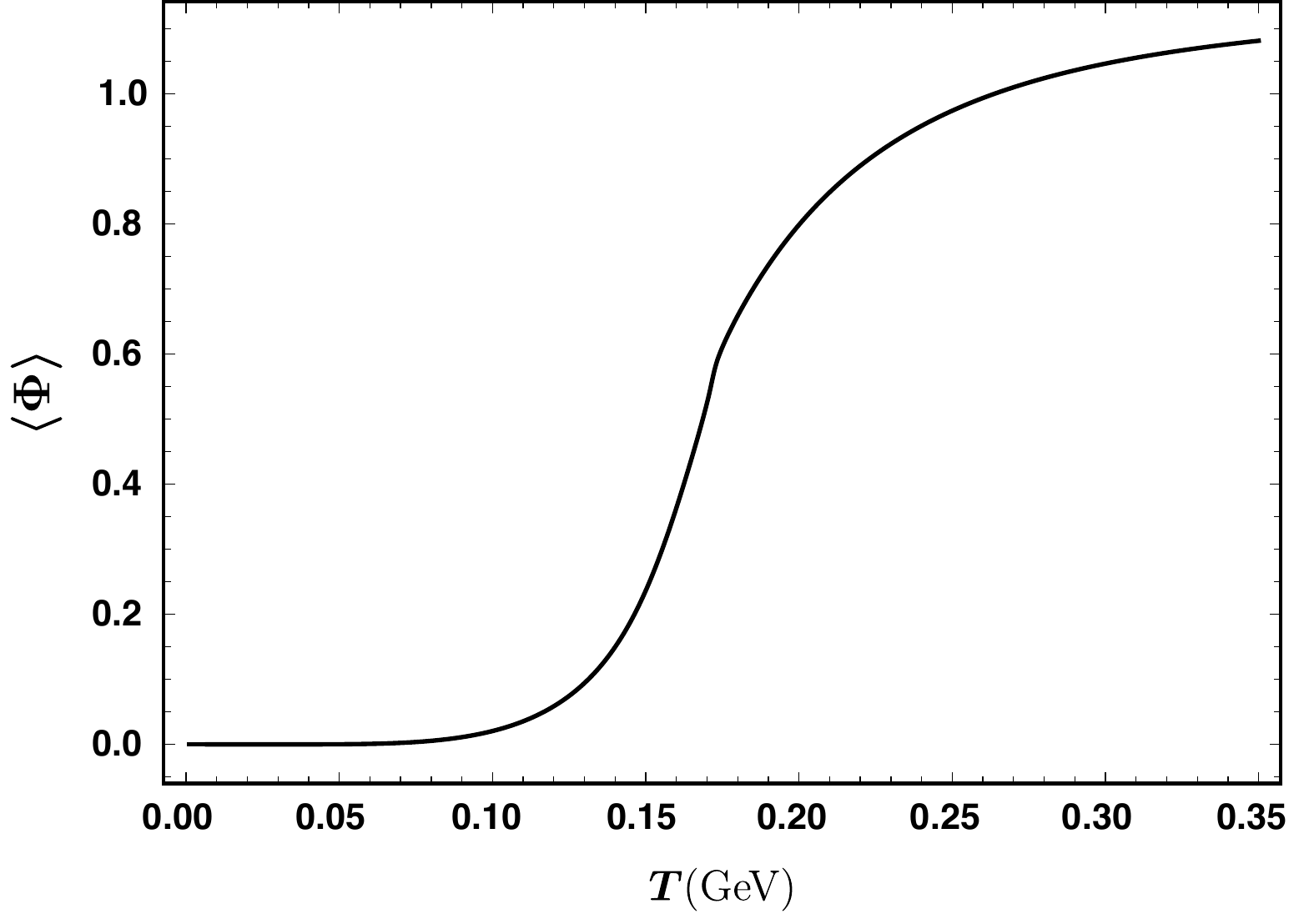}
 		\end{center}
 		\caption{Variation of $\sigma_x$, $\sigma_y $ condensate (left panel ) and the expectation value of Polyakov loop $\langle \Phi \rangle$ (right panel) as functions of temperature for $\mu_f=0$}
 		\label{order_para}
 	\end{figure}
 \end{center}
By solving the coupled equations in \eqref{gap_eq} one can extract the temperature dependence of strange condensate ($\sigma_y$), nonstrange condensate $(\sigma_x)$ and also the expectation value of the Polyakov loop.
The temperature variation of  $\sigma_x$ and $\sigma_y$ are plotted in the left panel and the expectation value of Polyakov loop $\langle \Phi \rangle$ is shown in the right panel of Fig.~\ref{order_para} at zero chemical potential ($\mu=0$) in PQM model. The quark condensates  $\sigma_x$ and $\sigma_y$ are regarded as order parameters of chiral phase transition in $(2+1)$ flavor PQM model. From the left panel of Fig.~\ref{order_para} it is shown that chiral symmetry is restored at high temperature. On the other hand $\langle \Phi \rangle$ increases from $\langle \Phi \rangle=0$ at $T=0$ to about  $\langle \Phi \rangle=1$ at high temperature. The detailed discussion can be found in Ref.~\cite{Schaefer:2009ui, Schaefer:2008hk}.

\begin{center}
 	\begin{figure}[tbh]
 		\begin{center}
 			\includegraphics[scale=0.5]{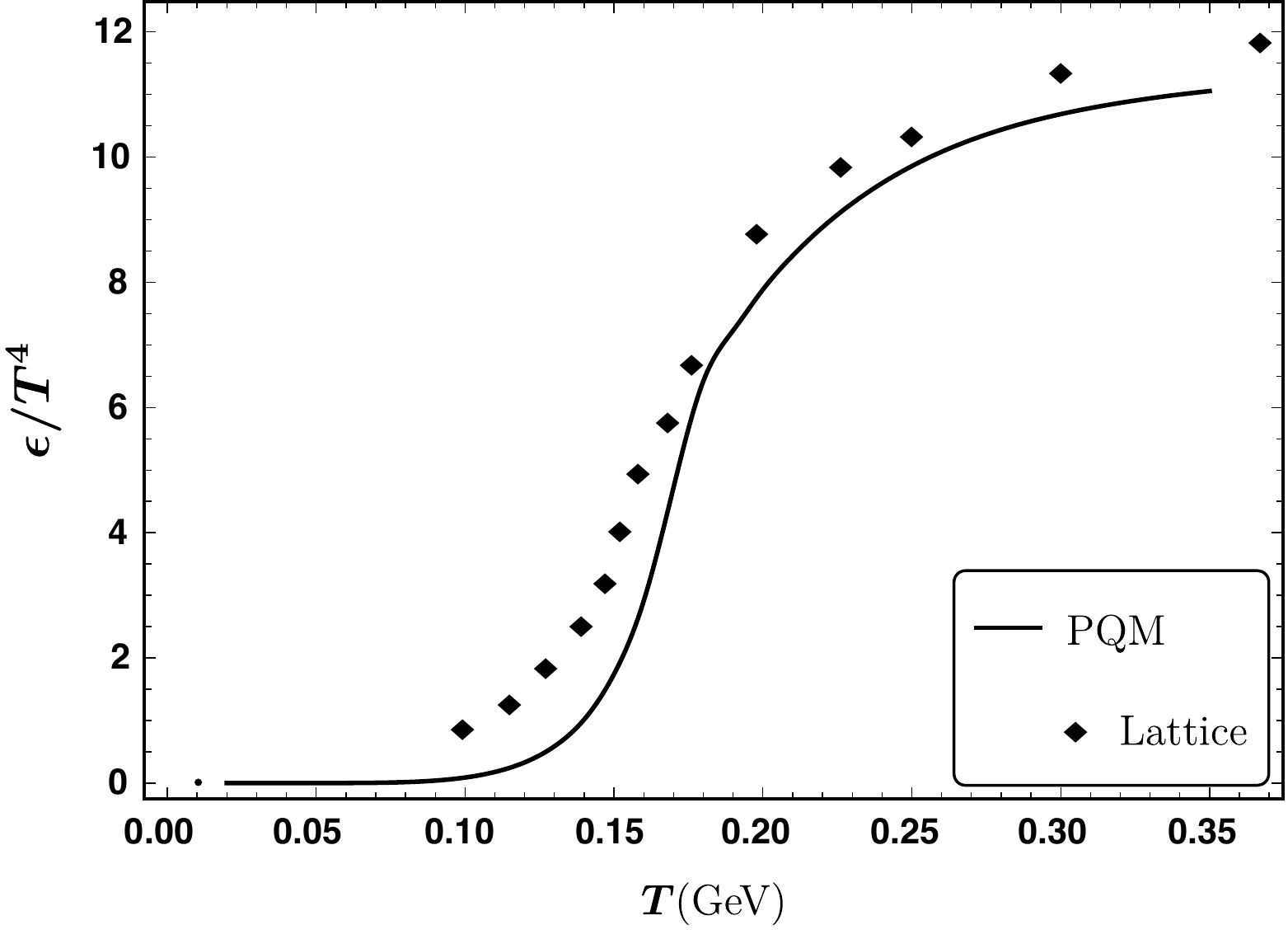}
 			\includegraphics[scale=0.5]{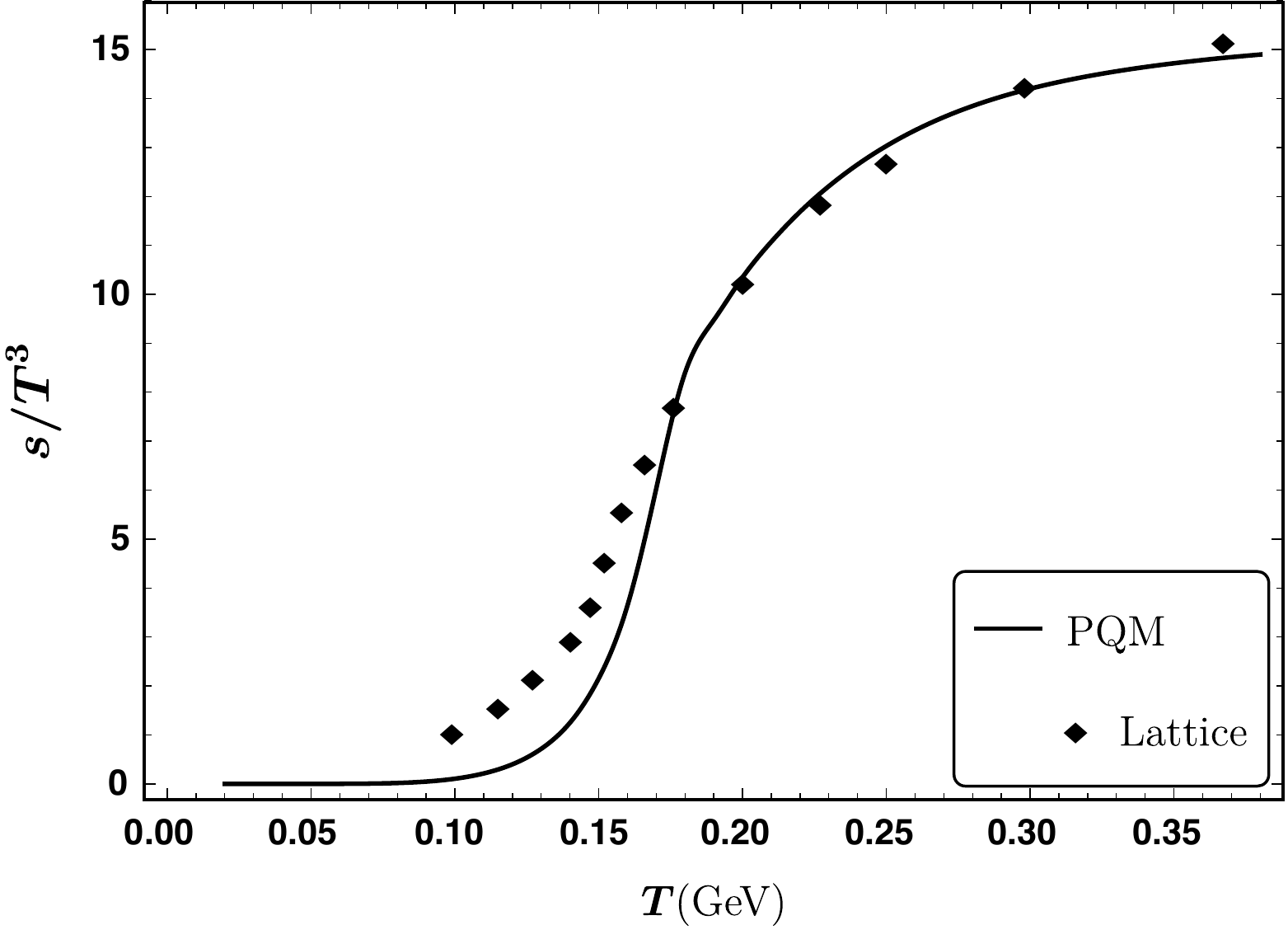}
 			\includegraphics[scale=0.5]{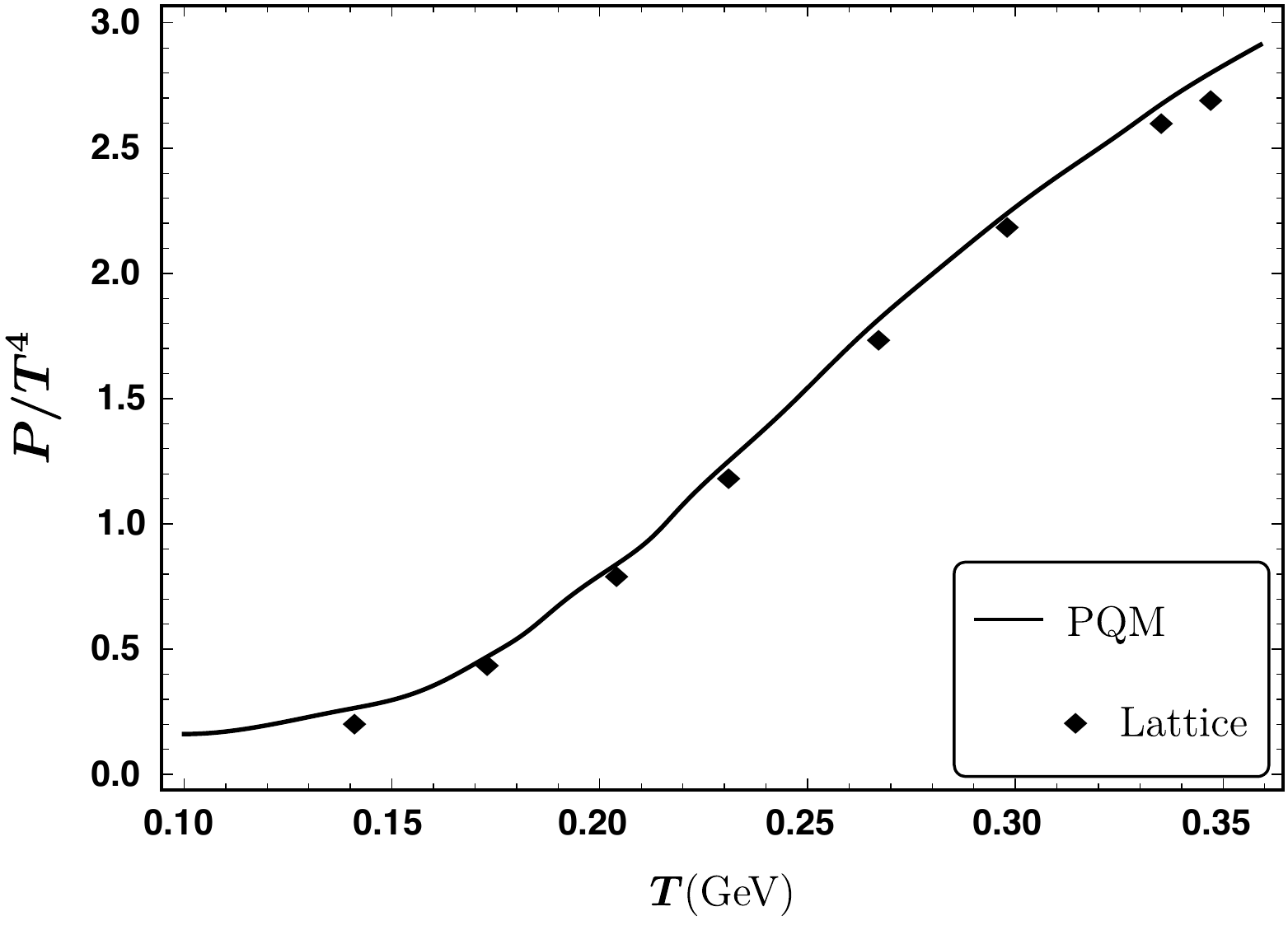}
 		\end{center}
 		\caption{Energy density (left), entropy density (right) and pressure (bottom) as functions of temperature for $\mu_f=0$}
 		\label{thermo}
 	\end{figure}
 \end{center}

 We have studied the thermodynamic quantities energy, pressure and entropy density. Pressure density is defined as 
 \bea
 P(T)=-\Omega(T).
 \eea
 By differentiating the thermodynamic potential, other quantities are obtained. The entropy density(s) and energy density $(\eps)$ are defined as
 \bea
 s(T)&=&-\frac{\partial \Omega(T)}{\partial T}
 \eea
 and 
 \bea
 \eps(T)=-P(T)+T s(T).
 \eea
 At high temperature, they show the behavior $P, \eps \sim T^4$ and $s\sim T^3$.
 In Fig.~\ref{thermo}, the scaled quantities $s/T^3, P/T^4$ and $\eps/T^4$ are plotted with the variation of temperature. We have plotted the graphs for zero chemical potential. They grow with the temperature and tend to saturate at high temperature. We have also compared the thermodynamic quantities with the LQCD results from Ref.~\cite{Borsanyi:2010cj}. The lattice data is shown for temporal extent $N_t=10$.  We found our results are in agreement with corresponding LQCD results.
 Here we want to mention that, implicit dependence of the order parameter on temperature as well as chemical potential should be taken into account to calculate various thermodynamic quantities. After solving the Eq.~\ref{gap_eq}, one can get the thermodynamic quantities by numerical differentiation of of the order parameters. To avoid the inaccuracy in the results, semianalytic approach should be taken~\cite{Ghosh:2014zra}.
 
\section{Transport coefficients }
\label{transport}
We have calculated shear viscosity coefficients for quark matter in relaxation time approximation. Medium dependent particle masses coming from the PQM model are considered. The viscosity coefficients are estimated for finite chemical potential.  The shear viscosity coefficient is given by~\cite{Deb:2016myz,Marty:2013ita}
\bea
\eta&=& \frac{1}{15T} \sum_a \int \frac{d^3p}{(2\pi)^3} \frac{p^4}{E_a^2} \tau f_a^0 (1-f_a^0),
\label{eta}
\eea
where $f_a^0$ is equilibrium Fermi-Dirac distribution function and $\tau_a$ is the relaxation time for particle `a'. The sum is over the three quark species including their antiparticles. The relaxation time is calculated from elastic scattering of quarks and antiquarks via gluon exchange in perturbative QCD.
   
   \subsection{Relaxation time estimation from elastic scattering in perturbative QCD}
Thermal relaxation time for a single species can be written as~\cite{Rehberg:1996vd, DeGroot:1980dk}
\bea
\tau_f^{-1}=\sum_{f'}\rho_{f'} \bar w_{ff'},
\eea   
where $\bar w_{ff'}$ is the sum of transition rates of all species.The averaged transition rate is defined as 
\bea
\bar w=\int_{\text{Th}}^{\infty} ds \sqrt{(p_1p_2)^2-(m_1m_2)^2}\sigma(s)P(s,T,\mu),
\eea
with threshold $\text{Th}=Max[(m_1+m_2)^2,(m_3+m_4)^2]$.
The weight function is written as 
\bea
P(s,T,\mu)&=& \frac{1}{\rho_1(T,\mu)\rho_2(T,\mu)}\frac{1}{16\pi^4}\int_{m_1}^{\infty}dE_1 f^0(\beta E_1,\mu) \int_{m_2}^{\infty}dE_2 f^0(\beta E_2,\mu)\nn\\
&\times& \Theta\bigg(4|\vec p_1|^2|\vec p_2|^2-(s-m_1^2-m_2^2-2E_1E_2)^2\bigg),
\eea
 where the quark density $\rho_f$ is defined as 
 \bea
 \rho_f(T,\mu)=\int \frac{d^3p}{(2\pi)^3} f^0\bigg(\beta\sqrt{\vec p^2+m_f^2},\mu\bigg),
 \eea 
 with $\beta=1/T$.
Now we need to calculate scattering cross section $\sigma(s,T)$ to obtain the relaxation time. To be specific we confine ourselves by taking the contribution from the elastic scattering of quarks and antiquarks via gluon exchange.
Including the fermi blocking factors, the total cross section in terms of differential cross section  looks like~\cite{Zhuang:1995uf} 
\bea
\sigma(s)=\int dt \frac{d\sigma}{dt}\bigg[1-f^0(\beta E_3,\mu)\bigg]\bigg[1-f^0(\beta E_4,\mu)\bigg],
\eea
where $E_{3,4}^2=p_{3,4}^2+m_{3,4}^2$.

For quark-quark scattering quarks can interact through $t$ and $u$ channel. There are four independent processes out of total six different possibilities for quark-quark scattering~\cite{Rehberg:1996vd}. 
All six processes are
\bea
uu\rightarrow uu ,\,\,\, && ss\rightarrow ss ,\,\,\, dd\rightarrow dd, \nn\\
ud\rightarrow ud ,\,\,\, && us\rightarrow us ,\,\,\, ds \rightarrow ds.
\eea

The differential cross section for this case is~\cite{Rehberg:1996vd,Cutler:1977qm} 
\bea
\frac{d\sigma}{dt}(q_\al q_\beta\rightarrow q_\al q_\beta)&=& \frac{1}{16\pi}\frac{1}{[s-(m_\al-m_\beta)^2][s-(m_\al+m_\beta)^2]}\frac{1}{36}\sum_{spin, color}|\mathcal{M}_t+\delta_{\al\beta}\mathcal{M}_u|^2,
\eea
with color and spin summed square of matrix element 
\bea
\frac{1}{36}\sum_{spin, color}|\mathcal{M}_t+\delta_{ff'}\mathcal{M}_u|^2&=&\frac{64\pi^2 \al_s^2}{9}\bigg[\frac{2(s-m_\al^2-m_\beta^2)^2+t^2+2st}{(t-m_g^2)^2}+\delta_{\al\beta}\frac{2(s-2m_\al^2)^2+u^2+2su}{(u-m_g^2)^2}\nn\\
&-&\delta_{\al\beta}\frac{2}{3}\frac{(s-4m_\al^2)^2-4m_\al^2}{(t-m_g^2)(u-m_g^2)}\bigg].
\eea
$\mathcal{M}_t$ and $\mathcal{M}_u$ are matrix elements for $t$ and $u$ channel respectively.

For quark-antiquark scattering, there are fifteen possible interaction processes. But from these fifteen possibilities, one get only seven independent processes.
Here all the fifteen possibilities are
\bea
&&u \bar d \rightarrow u\bar d ,\,\,\,\, d \bar u \rightarrow d\bar u ,\,\,\,\, s \bar u \rightarrow s\bar u,\nn\\
&&u \bar s \rightarrow u\bar s ,\,\,\,\, d \bar d \rightarrow d\bar s ,\,\,\,\, s \bar d \rightarrow s\bar d,\nn\\
&&u \bar u \rightarrow u\bar u ,\,\,\,\, d \bar d \rightarrow d\bar s ,\,\,\,\, s \bar d \rightarrow d\bar d,\nn\\
&&u \bar u \rightarrow d\bar d ,\,\,\,\, d \bar d \rightarrow s\bar s  ,\,\,\,\, s \bar s \rightarrow u\bar u,\nn\\
&&u \bar u \rightarrow s\bar s ,\,\,\,\, d \bar d \rightarrow u\bar u ,\,\,\,\, s \bar s \rightarrow s\bar s.
\eea
Quark and antiquark interact through $s$ and $t$ channel and the differential cross section is given by~\cite{Rehberg:1996vd,Cutler:1977qm} 
\bea
\frac{d\sigma}{dt}(q_\al \bar q_\beta\rightarrow q_\delta \bar q_\gm)&=&\frac{1}{16\pi}\frac{1}{[s-(m_\al-m_\beta)^2][s-(m_\al+m_\beta)^2]}\nn\\
&\times& \frac{64\pi^2\al_s^2}{9}\bigg[\delta_{\al\beta}\delta_{\gm\delta}\frac{2(u-m_\al^2-m_\delta^2)^2+s^2+2su}{(s-m_g^2)^2}\nn\\
&+&\delta_{\al\delta}\delta_{\beta\gm}\frac{2(u-m_\al^2-m_\beta^2)^2+t^2+2tu}{(t-m_g^2)^2}-\frac{2}{3}\delta_{\al\beta}\delta_{\beta_\delta}\delta_{\gm\delta}\frac{(u-4m_\al^2)^2-4m_\al^4}{(s-m_g^2)(t-m_g^2)}\bigg].
\eea
$\alpha_s$ is QCD coupling strength. To avoid the infrared divergent, effective gluon mass,
\bea
m_g^2=2\pi\al_s (1+\frac{1}{6}N_f)T^2,
\eea
is used as a regulator~\cite{Shuryak:1980tp, Rehberg:1996vd}.
   
 \begin{center}
	\begin{figure}[tbh]
		\begin{center}
			\includegraphics[scale=0.6]{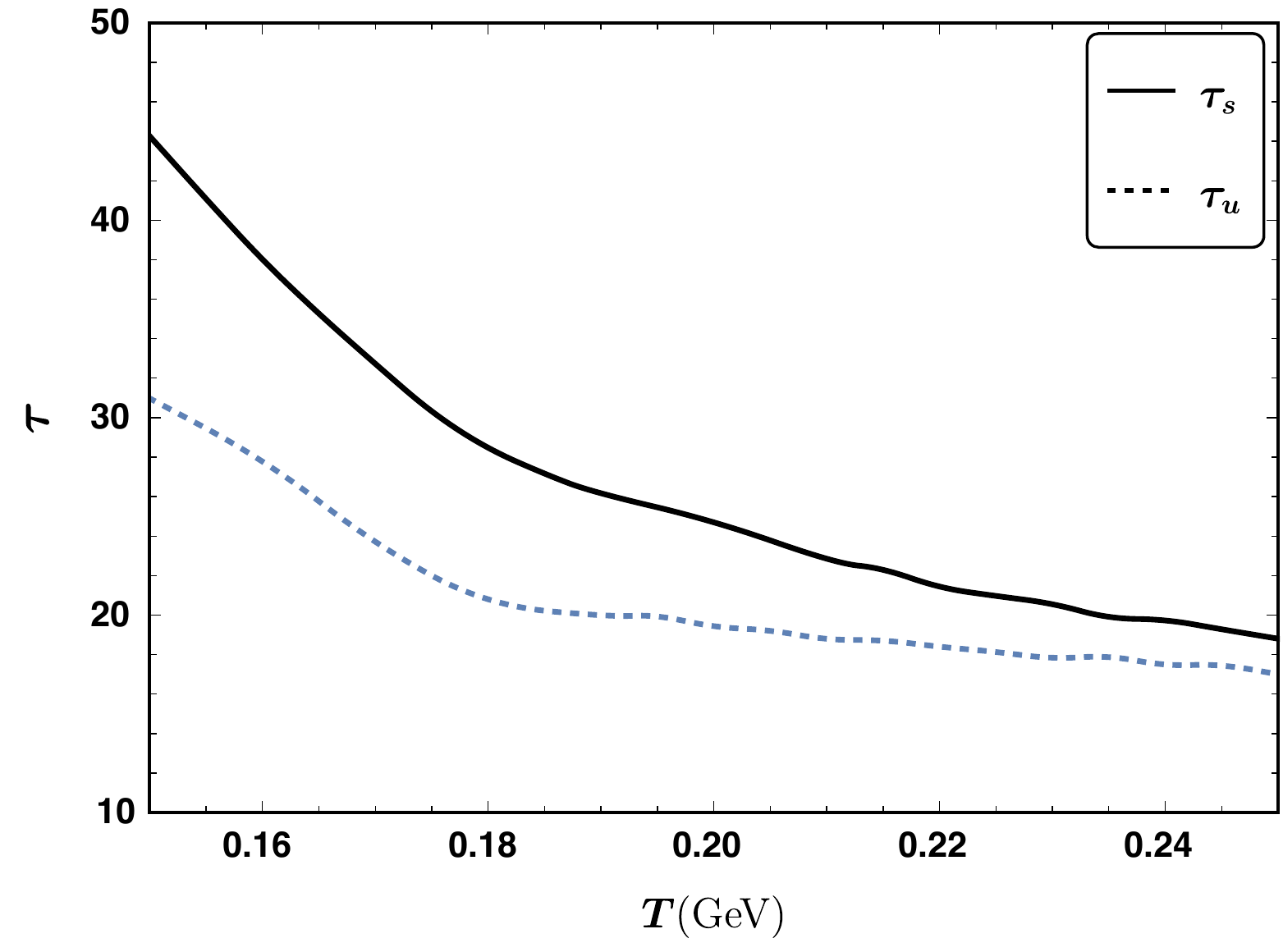}
		\end{center}
		\caption{Plot of thermal relaxation time for light and strange quark as a function of temperature.}
		\label{relaxation}
	\end{figure}
\end{center}

\section{Results}

\label{result}
 \begin{center}
	\begin{figure}[tbh]
		\begin{center}
			\includegraphics[scale=0.6]{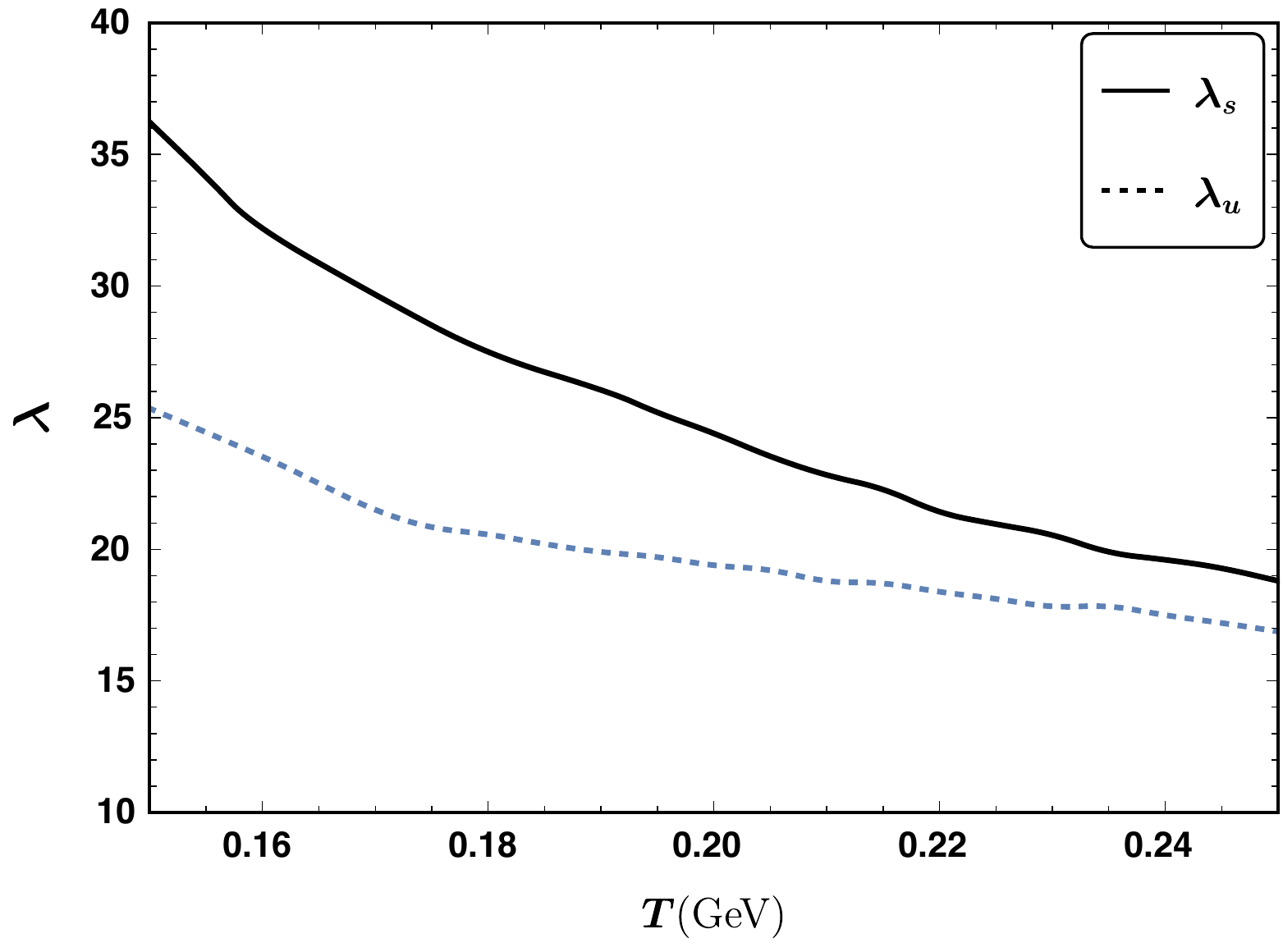}
		\end{center}
		\caption{Plot of mean free path for light and strange quark as a function of temperature}
		\label{mean_free}
	\end{figure}
\end{center}

Relaxation time is calculated numerically for quarks. Variation of relaxation times for up quark ($\tau_u$) and strange quark $(\tau_s)$ with temperature are shown in Fig.~\ref{relaxation}. $\tau_d$ for down quark is same as $\tau_u$. As quark density is low at higher temperature, relaxation time is large in this temperature region. $\tau_s$ is greater than $\tau_u$. From Fig.~\ref{order_para} it is shown that the condensate $\sigma_y$ is greater than $\sigma_x$. Now the effective masses of strange and nonstrange quarks are related via Eq~\eqref{quark_mass}. So it is clear that the effective mass of strange quarks are larger than the light quarks. Due to the larger effective mass, strange quark would take more time to relax than lighter quark. It should be noted that the greater temperature region than the Mott temperature is physically interested.

Here we define mean free path of particle. It is related to the relaxation time for each flavor as
\bea
\lambda_f&=& \tau_f \frac{6}{\rho_f} \int \frac{d^3p}{(2\pi)^3} \frac{p}{E_f} f^0(E_f/T).
\label{lambda}
\eea
The term multiplied with the relaxation time in eq.~\eqref{lambda} is the mean velocity. In fig.~\ref{mean_free}, mean free paths of quark flavors are shown. As $\tau_s$ is greater the $\tau_u$, it is expected that $\lambda_s$ is also higher than $\lambda_u$.

 \begin{center}
	\begin{figure}[tbh]
		\begin{center}
			\includegraphics[scale=0.6]{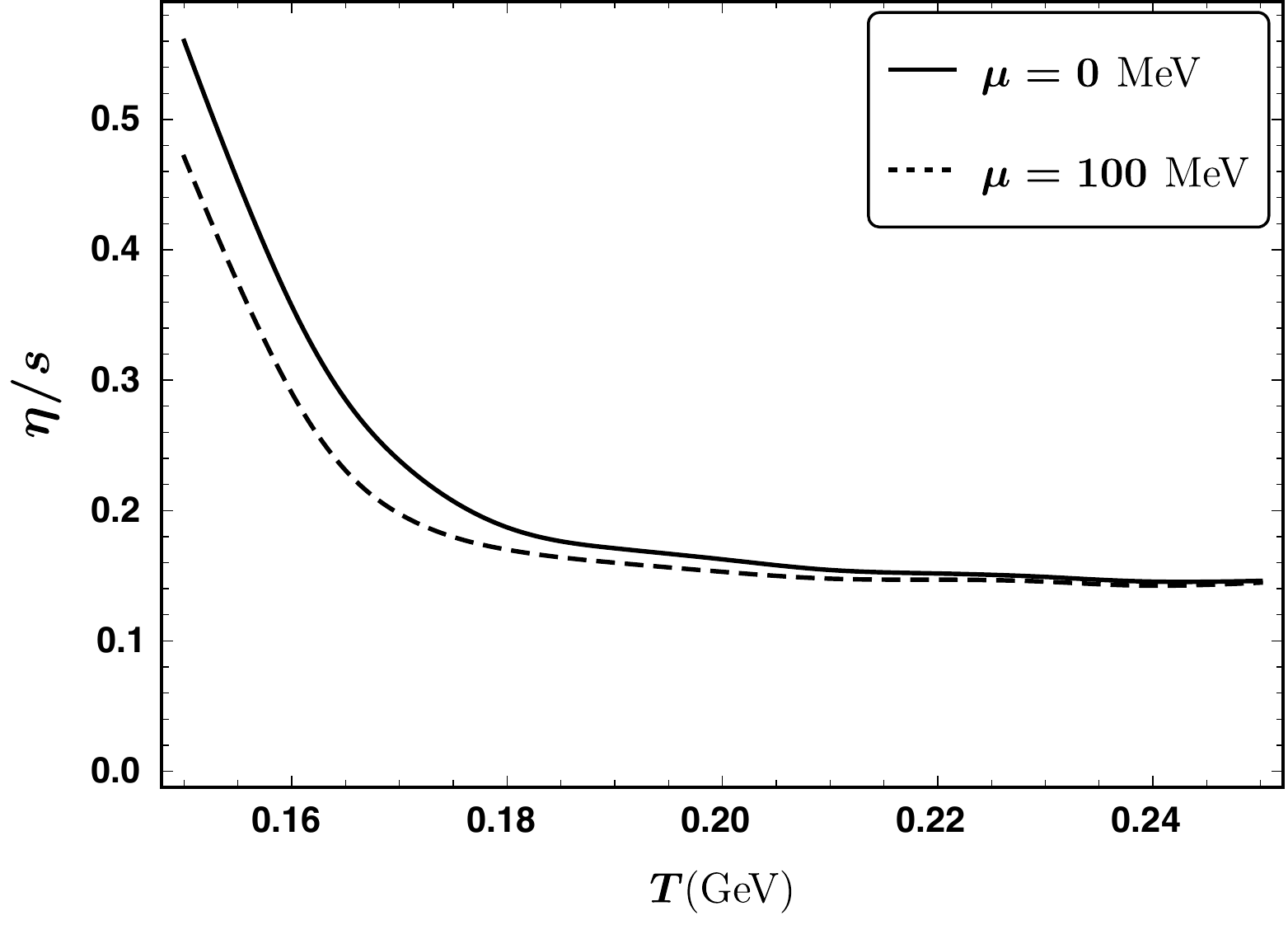}
		\end{center}
		\caption{viscosity over entropy ratio as a function of temperature for zero and finite chemical potential}
		\label{viscosity}
	\end{figure}
\end{center}

We have calculated shear viscosity numerically from eq.~\eqref{eta} using the relaxation time for different chemical potential.
In fig.~\ref{viscosity}, specific shear viscosity $(\eta/s)$ is plotted as a function of temperature $(T)$ for zero and finite chemical potential $(\mu=100 \,\,\text{MeV})$. $\eta/s$ falls rapidly near the transition temperature. We are not getting true minimum near critical temperature. One needs to include mesonic contribution to the relaxation time coming through the interaction. At finite chemical potential, $\eta/s $ is lower than the value of the same at vanishing $\mu$. Value of $\eta/s$ is well above the Kovtun-Son-Starinets (KSS) bound~\cite{Kovtun:2004de} of $1/4\pi$.

\section{Summary}
\label{summary}
In present study, we have attempted to investigate the temperature behavior of shear viscosity coefficients which is one relevant input for hydrodynamic evolution of hot and dense QCD medium. The viscosity coefficient is estimated within the Polyakov loop extended quark meson model for three quark flavors. The approach uses the relaxation time approximation of the Boltzmann equation where the particle masses are medium dependent. The expression of the shear viscosity coefficient is positive definite as it should be. 

Medium dependent quark masses are evaluated within mean field approximation. QCD thermodynamics of PQM model for three flavors ($u,d,s$) is also discussed. The thermodynamic quantities are also compare with lattice result for temporal extent $N_t=10$ .The relaxation time for strange and non-strange quarks are obtained from the quark quark and quark-antiquark scattering. The mean free paths for both types of quarks are also discussed. Relaxation time as well as the mean free path for strange particle is higher than the non-strange particle.  In this study we are limited only in quark scatterings. The shear viscosity coefficient is decreasing rapidly near critical temperature. In future a more rigorous study should be done by considering also the quark meson scatterings to get more improved results of viscosity coefficients.

\section{Acknowledgement}
RG is funded by University Grants Commission (UGC). The author acknowledges  Munshi G. Mustafa for immense encouragement and support. The author also like to thank Arghya Mukherjee for helpful discussions.

\bibliographystyle{apsrev4-1}
\bibliography{draft_pqm,revtex-custom}

\end{document}